\documentclass[twocolumn,showpacs,preprintnumbers,amsmath,amssymb]{revtex4}
\usepackage{amssymb}
\usepackage[dvips]{graphicx}
\begin{document}

\textbf{How to map a pseudogap?} 

A  pseudogap (PG) is believed to be responsible for the non Fermi-liquid normal state of cuprate superconductors. In particular, field induced PG collapse causes negative longitudinal magnetoresistance (MR) \cite{we}. The PG collapses because of spin-splitting of the polaron band while the orbital effects are irrelevant \cite{we}. Recently these conclusions, including the Zeeman relation, $k_BT^*=gB_{pg}$, which couples the PG temperature, $T^*$, and the PG closing field, $B_{pg}$, were reaffirmed by the authors of Ref.\cite{lia2}. It will be demonstrated below that \cite{lia2} lacks consistency and its conclusions are based on fallacious propositions and  unsupported by the authors' own experimental results. For these reasons,\cite{lia2} does not represent reliable independent evidence in support of the original findings by \cite{we}. 

Unlike \cite{we}, Ref.\cite{lia2} mistakenly assumes that it is not the conductance but the net resistance, $\rho_c=\rho^n+\Delta\rho_c$, that is the sum of two channels: the `ungapped' $\rho^n\propto T$, measured with intrinsic tunnelling at the highest overheating \cite{myPhysicaC}, and $\Delta\rho_c$, the excess resistivity due to DOS depletion \cite{fault} obtained by the subtraction of that poorly defined $\rho^n$ from the net $\rho_c(T,B)$. Ref.\cite{lia2} further claims that $\Delta\rho_c(B)$ extrapolation  beyond 60 T gives a reliable $B_{pg}$-estimate that is insensitive to the functional form of the fit, so that other approximations give the same estimates. I will show below that, in addition to the inconsistency [4] and the lack of theoretical support, the entirely empirical $B_{pg}$ evaluation procedure of \cite{lia2} lacks both reliability and accuracy. 

Providing the data from \cite{lia2} are reliable, these should allow for a cross-check of the authors' conclusions. However, even a brief look at the insert to Fig.1c reveals several inconsistencies. First, contrary to \cite{lia2}, the power-law fit here appears to be a 3rd order polinomial. Second, unlike \cite{lia2}, which claims $B_{pg}=300\pm50T$, I found that extrapolations of data-compatible fits give $B_{pg}$ in the range 200-$\infty$ for the same crystal, UD($T_c$=90K). Thus, the accuracy claimed by \cite{lia2} is seriously overestimated. 

Moreover, the data of the most overdoped sample, central to \cite{lia2}, are even more dubious.  Let us consider the insert to Fig.2 from \cite{lia2} (reproduced in Fig.1) which allegedly justifies both the power-law dependence, $\Delta\rho_c(B)-\Delta\rho_c(0)\propto B^\alpha$, and the accuracy of $B_{pg}$ estimate. As is clear from Fig.1, the experimental curves favour an exponential dependence (solid lines) that provides a reasonable T-dependent parameter $B_0$ (see Table in Fig.1) in drastic contrast to the unphysical scatter of $\alpha$ and $\Delta\rho_c(0)$, the parameters of the fit by \cite{lia2}. Importantly, Fig.1 suggests a dramatically higher uncertainty in the extrapolation procedure by \cite{lia2}. Even if this procedure is adequate, a realistic $B_{pg}$ estimate from Fig.1 gives 100-1000T rather than 86-89 as in \cite{lia2}. Thus, the experimental data of \cite{lia2} do not support their conclusions.

\begin{figure}
\begin{center}
\includegraphics[angle=-0,width=0.47\textwidth]{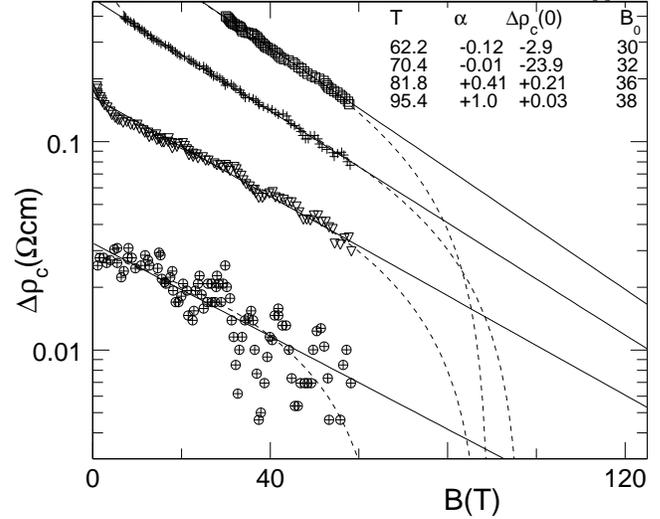}
\vskip -0.5mm 
\caption{$\Delta\rho_c(B)$ by \cite{lia2} (symbols) are fitted to the dependence proposed in \cite{lia2}, $\Delta\rho_c(B)-\Delta\rho_c(0)\propto B^\alpha$, and to $\Delta\rho_c(B)\propto exp(-B/B_0)$, which are shown by the broken and solid lines respectively; the table shows the fitting parameters obtained. }
\end{center}
\end{figure}

To conclude, the  $B_{pg}$ evaluation procedure by  \cite{lia2} is theoretically unjustified and could only be used as an empirical exercise. Moreover, the experimental data by \cite{lia2} do not provide the accuracy claimed  for the $B_{pg}$ estimate, thus rendering irrelevant their phase diagrams. Hence the conclusions of \cite{lia2} lack reliable grounds beyond those of Ref.\cite{we}. Additional inconsistencies in \cite{lia2} carried over from prior articles [4] cast  further doubts on the reliability of \cite{lia2}. However, as far as the raw data are concerned, these may not necessarily be incorrect. In particular, the estimates of resistive upper critical field correlate reasonably with \cite{we}. Interestingly, $H_{c2}$ for crystals of vastly different doping \cite{lia2} follow the single dependence, $H_{c2}=H_0(t^{-1}-t^{1/2})^{3/2}$ ($t=T/T_{c}$), with the doping dependent $H_0$=4-12T, see \cite{my_jetp} for details. 

I am grateful to Nai-Chang Yeh for drawing my attention to \cite{lia2}, to A. Alexandrov for stimulating discussions and to the Leverhulme Trust for financial support.

V.N.Zavaritsky 

Department of Physics, Loughborough University, Loughborough LE11 3TU, United Kingdom.


\end{document}